\documentclass[twocolumn,twoside,slac_two]{revtex4}
\usepackage{graphicx}
\usepackage{fancyhdr}
\pdfoutput=1

\begin{document}

\title{Radiation from Particles Accelerated in Relativistic Jet Shocks and Shear-flows}

%

\author{K.-I. Nishikawa}
\affiliation{Department of Physics, University of Alabama in Huntsville, ZP12, Huntsville, AL 35899, USA}
\author{P. Hardee}
\affiliation{Department of Physics and Astronomy, The University of Alabama, Tuscaloosa, AL 35487, USA}
\author{I. Dutan }
\affiliation{Institute of Space Science, Atomistilor 409, Bucharest-Magurele RO-077125, Romania}
\author{B. Zhang}
\affiliation{Department of Physics, University of Nevada, Las Vegas, NV 89154, USA}
\author{A. Meli}
\affiliation{Department of Physics and Astronomy, University of Gent, Proeftuinstraat 86 B-9000, Gent, Belgium}
\author{E. J. Choi, and K. Min}
\affiliation{Korea Advanced Institute of Science and Technology, Daejeon 305-701, South Korea}
\author{J. Niemiec}
\affiliation{Institute of Nuclear Physics PAN, ul. Radzikowskiego 152, 31-342 Krak\'{o}w, Poland}
\author{Y. Mizuno}
\affiliation{Institute for Theoretical Physics, University of Frankfurt am Main, 60438, Germany}
\author{M. Medvedev}
\affiliation{Department of Physics and Astronomy, University of Kansas, KS 66045, USA}
\author{\AA \ Nordlund, and J. T. Frederiksen}
\affiliation{Niels Bohr Institute, University of Copenhagen, Juliane Maries Vej 30, 2100 Copenhagen Ø, Denmark}
\author{H. Sol}
\affiliation{LUTH, Observatore de Paris-Meudon, 5 place Jules Jansen, F-92195 Meudon Cedex, France}
\author{M. Pohl}
\affiliation{Institut fur Physik und Astronomie, Universit\"{a}at Potsdam, D-14476 Potsdam-Golm, Germany;
DESY, Platanenallee 6, D-15738 Zeuthen, Germany}
\author{D. Hartmann}
\affiliation{Department of Physics and Astronomy, Clemson University, Clemson, SC 29634, USA}

\begin{abstract}
We have investigated particle acceleration and emission from shocks and shear flows associated with an unmagnetized relativistic jet plasma propagating into an unmagnetized ambient plasma. Strong electro-magnetic fields are generated in the jet shock via the filamentation (Weibel)  instability. Shock field strength and structure depend on plasma composition (($e^{\pm}$ or $e^-$- $p^+$ plasmas) and Lorentz factor. In the velocity shear between jet and ambient plasmas, strong AC ($e^{\pm}$ plasmas) or DC ($e^-$- $p^+$ plasmas) magnetic fields are generated via the kinetic Kelvin-Helmholtz instability (kKHI), and the magnetic field structure also depends on the jet Lorentz factor. We have calculated, self-consistently, the radiation from electrons accelerated in shock generated magnetic fields. The spectra depend on the jet's initial Lorentz factor and temperature via the resulting particle acceleration and magnetic field generation.  Our ongoing ``Global" jet simulations containing shocks and velocity shears will provide us with the ability to calculate and model the complex time evolution and/or spectral structure observed from gamma-ray bursts, AGN jets, and supernova remnants.
\end{abstract}
\maketitle
\thispagestyle{fancy}

\section{Introduction}

\vspace{-0.4cm}
Blazars and Gamma-Ray Bursts (GRBs) are the most luminous phenomena in the universe. Despite extensive observational and theoretical programs, our understanding of the physics remains quite limited. There is broad consensus that both are powered by relativistic jets, which are directly imaged by interferometry in the case of blazars, and that the jets are launched and collimated mainly by magnetic forces e.g., \cite{Vlah03,mck06,komisetal07,tchmn08}.  However, there is uncertainty regarding details such as (1) magnetic versus kinetic domination, (2) rapid acceleration of particles to GeV and TeV energies, (3) location of highly variable gamma-ray emission and (4) source of seed photons if inverse Compton, (5) the scale of magnetic field turbulence in the radiation zone(s), and (6) the role of large and small scale instabilities in jet structure, dynamics, magnetization, particle energization and radiation. 

For blazars, the observational data are now quite rich, with dense time sampling of flux at many frequencies from radio to GeV and TeV $\gamma$-ray, linear polarization at radio to optical, and images with sub-parsec linear resolution at mm wavelengths  e.g., \cite{kra07,narapi12}. For GRBs, the basic measurements of $\gamma$-ray and X-ray flux vs.\ time during the burst have been supplemented by observations of the afterglow at soft X-ray, optical, IR, and radio frequencies. What is missing is a comprehensive theoretical framework for interpreting this wealth of observational data. There are numerous studies that consider only radiative processes, instabilities, or particle acceleration (see \cite{hP10}), and a smaller number that pair the first with one of the latter two e.g., \cite{dma12}. While separating analyses into soluble parts is a valuable technique, in GRB and AGN jets the dynamics, instabilities, and energy gains and losses are coupled processes. Here we present recent progress in shock and velocity shear simulations using a relativistic particle-in-cell ({\bf RPIC}) code. 

\vspace{-0.6cm}
\section{Shock Simulations}.pdf
\vspace{-0.35cm}
\subsection{Particle Acceleration and Magnetic Field Generation}
\vspace{-0.30cm}

{\bf RPIC} simulations have been used to study particle acceleration, magnetic field generation, and emission from collisionless shocks. Simulations reveal that the filamentation ({\bf Weibel}) instability, which generates magnetically wrapped current filaments, dominates relativistic shock processes \cite{diec06}.   There are significant differences between electron-positron pair and electron-ion shocks as the ion filamentation instability enhances shock magnetic field generation and thermal energy density relative to pair plasmas e.g., \cite{nhrpsf03,nishi05,nishi06,nishi07,nishi09a,fred03,fred04,hede05,ram07,Spitkovsky05,Spitkovsky08}.

Our 3-D MPI parallelized RPIC code has been used to simulate relativistic electron-ion jet propagation into an unmagnetized ambient electron-ion plasma ($m_{\rm i}/m_{\rm e} = 16$) with  equal jet and ambient electron number density, and jet thermal velocity $v^{\rm e}_{\rm j,th} = 0.2c$ where $c$ is the speed of light, and the jet Lorentz factor is $\gamma = 15$.  The simulation used a system with dimension ($L_{\rm x}, L_{\rm y}, L_{\rm z}) = (8192\Delta,$ $64\Delta, 64\Delta)$, where $\Delta$ is the cell size, and a total of $\sim 1$ billion particles (16 particles$/$cell$/$species for the ambient plasma) \cite{choi14}. This computational domain is twelve times longer than in our previous simulations (\cite{nishi06,ram07}). 
 
\begin{figure}[!h]
\includegraphics[width=75mm]{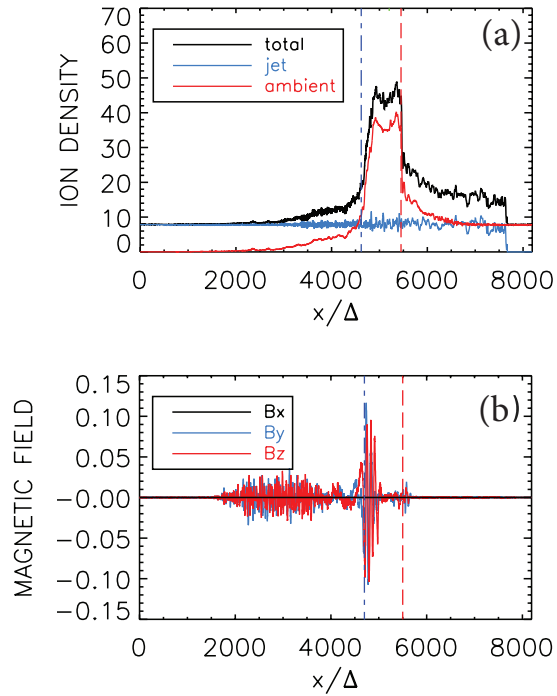}
\vspace*{-0.40cm}
\caption{\footnotesize \baselineskip 10pt Panel (a) shows the ion density (black: total, red: ambient, and blue: jet), and panel (b) shows the magnetic field components at the end of the simulation ($T =7372/\omega_{\rm pe}$). Electromagnetic field energy densities are normalized by the jet kinetic energy density, and quantities are averaged over the $yz-$plane.\label{weib}} 
\vspace{-0.2cm}
\end{figure}

Figure \ref{weib} shows the averaged (in the $y-z$ plane) ion density and electromagnetic field energy along the electron-ion jet at  simulation time $t = 7372\omega_{\rm pe}^{-1}$.  
%
The resulting profiles of jet (red), ambient (blue), and total (black) ion density are shown in Figure \ref{weib}a.  Warm jet ions are thermalized and ambient electrons are accelerated in the resulting leading (bow) and trailing (jet) shock system. Ambient ions are accelerated and pile up towards the jet front.  The ambient plasma density increases behind the jet front, with additional increase to a higher plateau farther behind the jet front indicating the leading shock. The jet ion density remains approximately constant. 
The {\bf strongest electromagnetic fields} are located at $x/\Delta =4,500$ as shown in Figure \ref{weib}b and are associated with the trailing shock. These strong fields may lead to the observed time dependent GRB afterglow emission. The longer simulation system has allowed significant non-linear shock Weibel instability and associated particle acceleration development. 

\vspace{-0.5cm}
\subsection{Self-Consistent Synthetic Spectra}
\vspace{-0.35cm}

We have calculated the radiation spectra directly from our simulations by integrating the expression for the retarded power, derived from Li\'{e}nard-Wiechert potentials, for a large number of representative particles in the PIC representation of the plasma 
\cite{jackson99,hedeT05,nishiP09b,nishi10a,nishiP10b,nishiP10c,nishi11,martins09,sironi09,fred10,medlet11}.

\begin{figure}[h]
\begin{center}
\includegraphics[width=43mm]{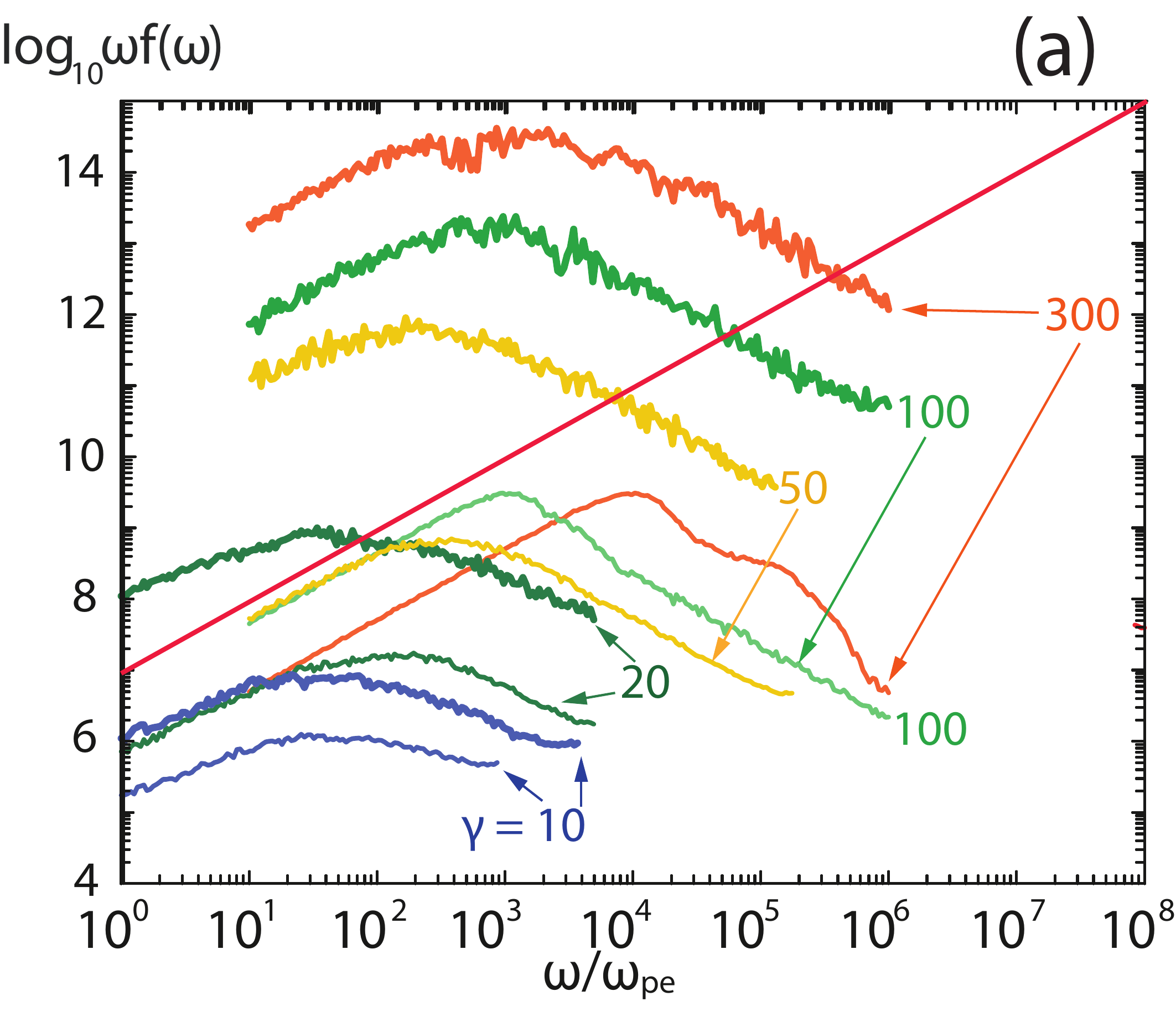}
\hspace*{-0.2cm}
\includegraphics[width=36mm]{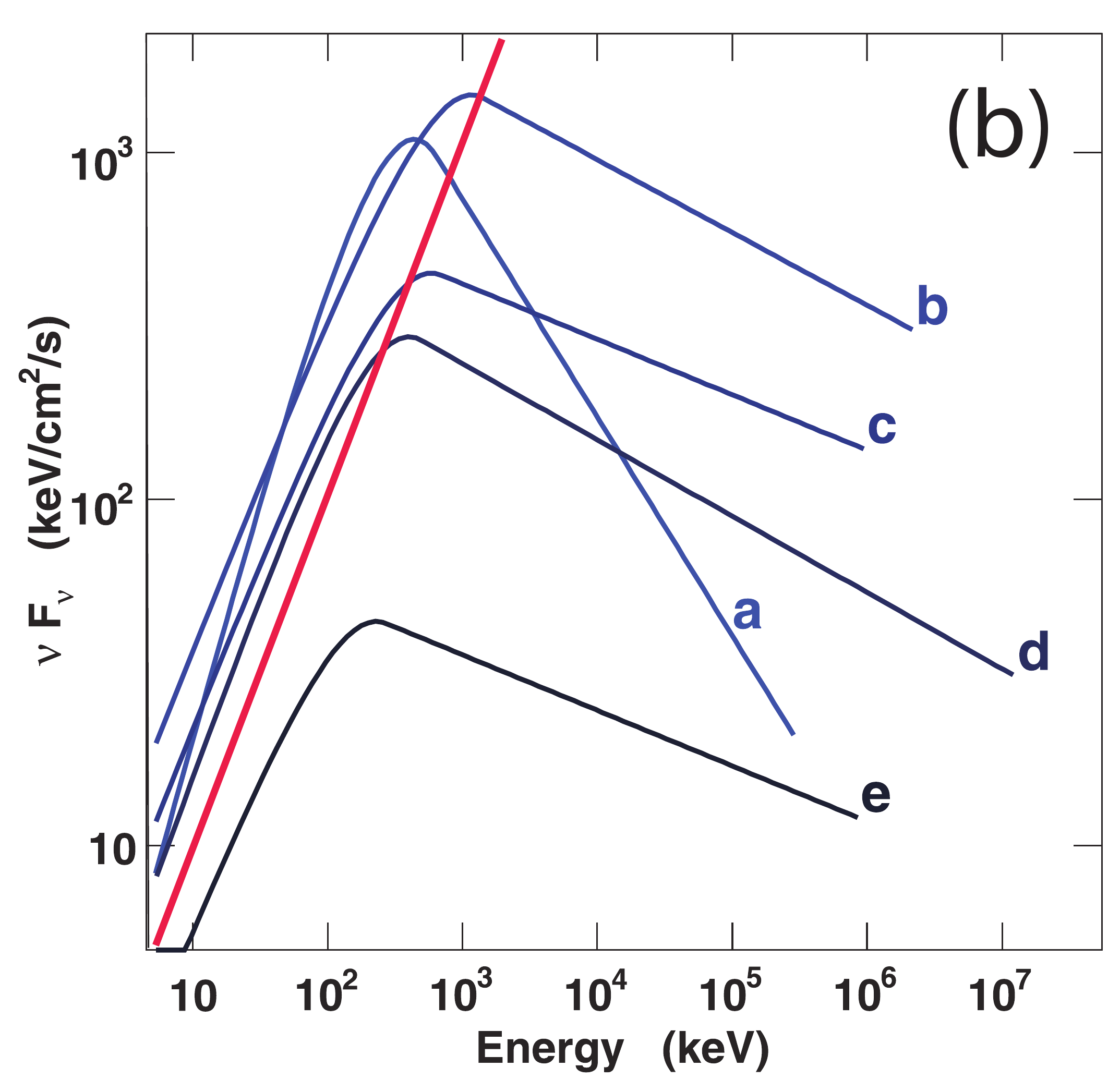}
\end{center}
\vspace*{-0.5cm}
\caption{\footnotesize \baselineskip 10pt  Panel (a) shows synthetic spectra for jets with Lorentz factors of $\gamma =$ 10, 20, 50, 100, and 300 and cold (thin lines) or warm (thick lines) jet electrons.    Panel (b) shows modeled Fermi spectra in $\nu F_{\nu}$ units at early (a) to late (e) times \cite{Abdo09}. Straight red lines indicate a slope of $\nu F_{\nu} = 1$. \label{spec}}
\vspace{-0.3cm}
\end{figure}

The synthetic spectra shown in Figure \ref{spec}a are obtained for emission from electrons in jets with {\bf Lorentz factors} of $\gamma =$ 10, 20, 50, 100, and 300. Spectra are obtained from an ensemble of electrons selected from the region where the Weibel instability, particle acceleration, and magnetic field generation are strongest.
\vspace*{-0.2cm}
For each jet Lorentz factor we computed spectra for cold, $v_{\rm jet,th} = 0.01c$, (thin lower lines) and warm, $v_{\rm jet,th} = 0.1c$, (thick upper lines) jet electrons \cite{nishiP11b,nishi12a,nishi12b,nishi12c}. Here the spectra are calculated for radiation beamed along the jet axis.  We note that radiation losses are not included \cite{jaro09,MS09}. 

Synthetic spectra are Bremsstrahlung-like at low frequencies (\cite{hedeT05})  because the magnetic fields generated by the Weibel instability are weak and electron acceleration is modest. Synthetic spectra low frequency slopes are very similar to those of the spectra shown in Fig. \ref{spec}b from \cite{Abdo09}. 
Comparison between our synthetic spectra and  the spectra from Abdo et al.\cite{Abdo09} suggest that the spectral evolution observed from early to late times is mimicked by our synthetic spectra evolution from higher to lower jet Lorentz factor. However, further investigation is necessary and this is one of our future research efforts.

\vspace*{-0.55cm}
\section{Velocity Shear Simulations}
\vspace*{-0.35cm}
\subsection{Slab Jet Velocity Shear}
\vspace*{-0.3cm}

\begin{figure}[h!]
\begin{minipage}[t]{45mm}
\hspace{-0.6cm}
\includegraphics[width=43mm]{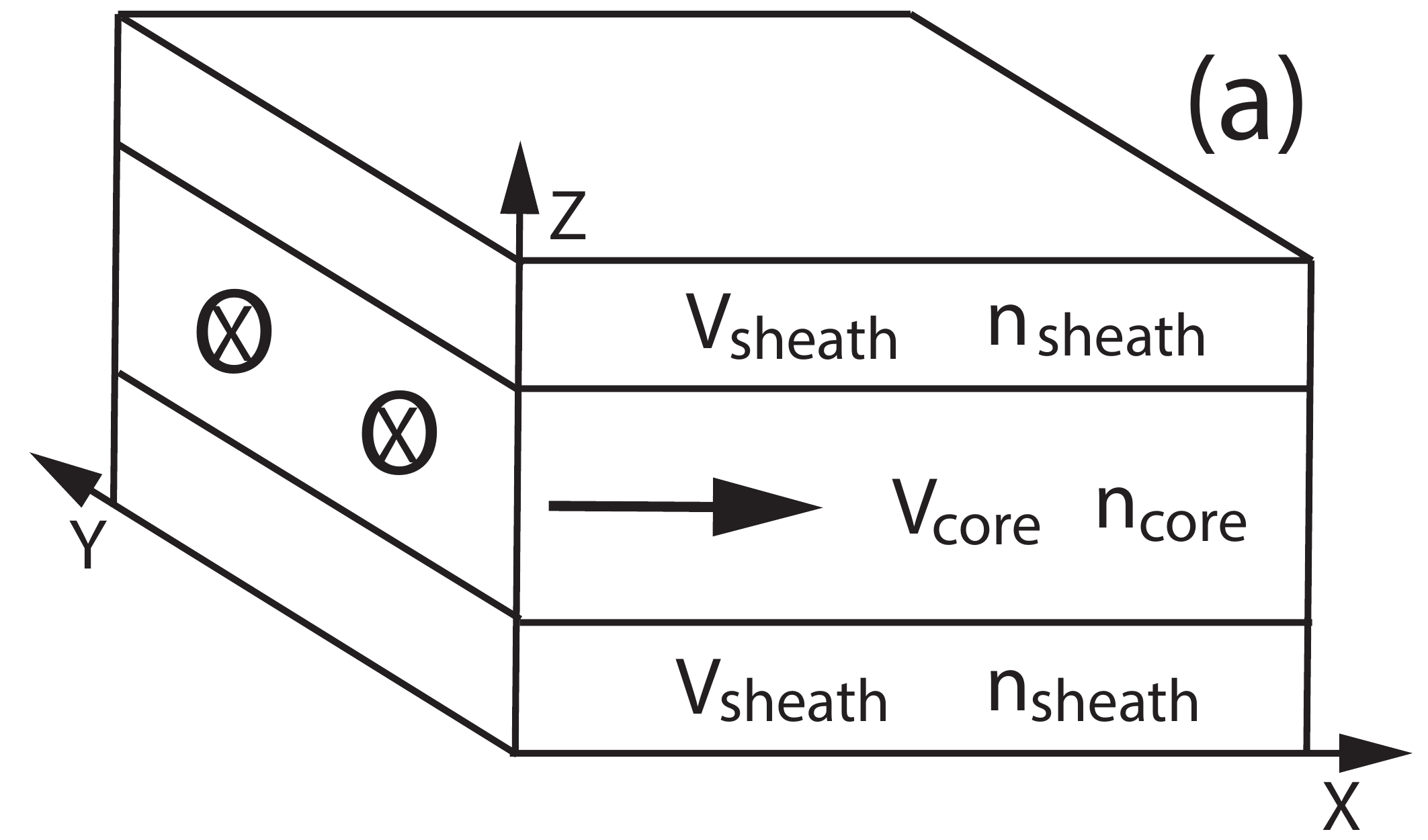}
\hspace*{-0.6cm}
\includegraphics[width=43mm]{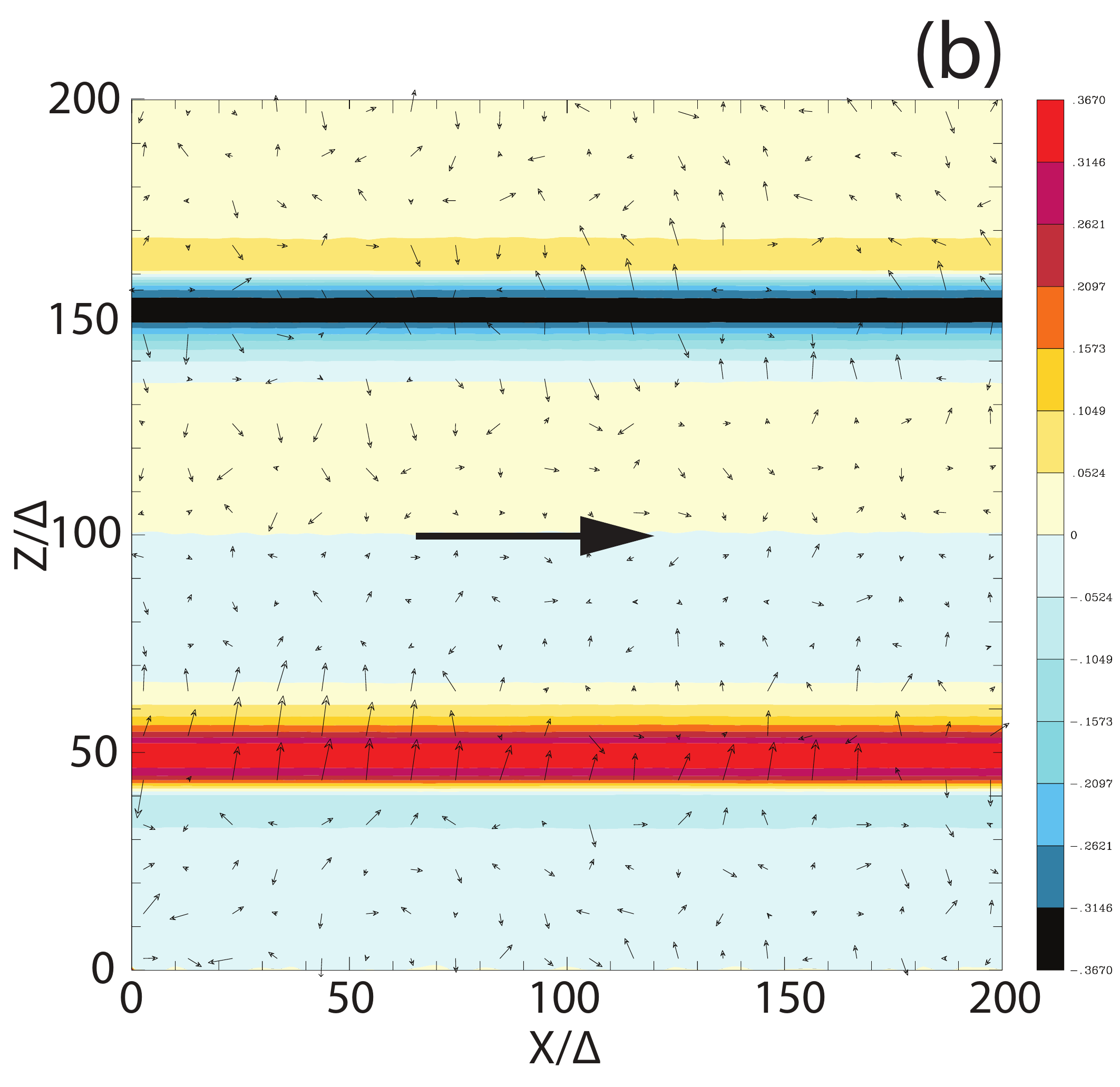}
\hspace*{-0.6cm}
\includegraphics[width=43mm]{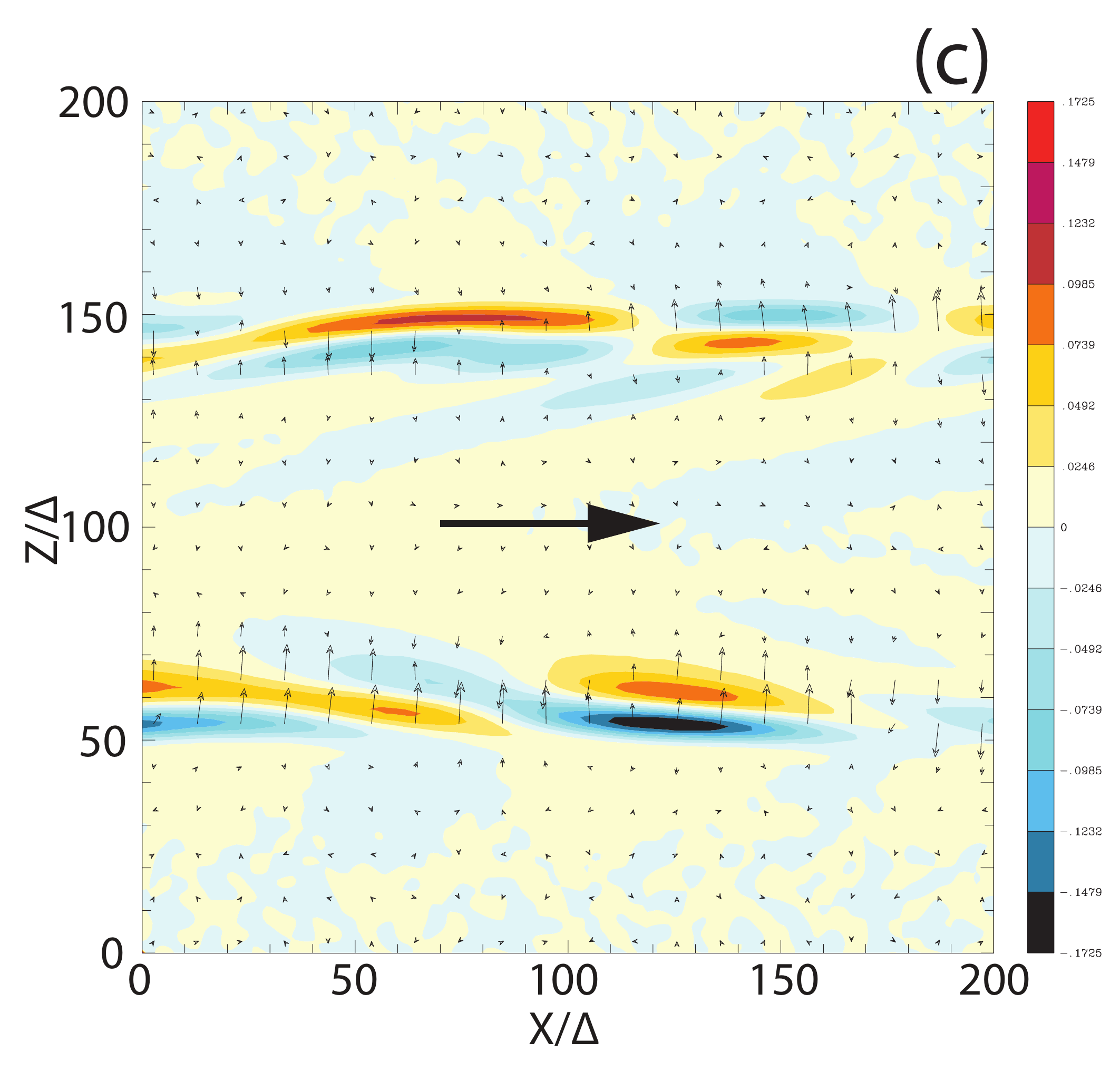}
\end{minipage}
\begin{minipage}[t]{30mm}
\vspace{-1.8cm}
\caption{\footnotesize \baselineskip 10pt  Panel (a)  shows our three-dimensional simulation setup. Panels (b) \& (c) show the magnetic field component $B_{\rm y}>0$ (red) and $B_{\rm y}<0$ (blue) plotted in the $x - z$ plane (jet flow indicated by large arrows) at the center of the simulation box, $y = 100\Delta$ at $t = 300\,\omega_{\rm pe}^{-1}$,  (b) for the $e^{-}$- $p^{+}$ case and (c) for the $e^{\pm}$ case, both with $\gamma_{\rm jt}=15$.  The smaller arrows indicate the magnetic field direction in the plane.   Panels (b) \& (c) cover one fifth of the simulation system length in the $x$ direction.  
 \label{setup}}
\end{minipage}
\end{figure}

In this simulation study we used a core-sheath plasma jet structure instead of the counter-streaming plasma setups used in previous simulations by \cite{Alves10,Alves12,Alves14,Gris13a,Gris13b,liang13a,liang13b}.
The basic setup and illustrative results are shown in Figure \ref{setup}.  
In our setup, a jet core with velocity $v_{\rm core}$ in the positive $x$ direction resides in the middle of the computational box (see Figure \ref{setup}a).  The  upper and lower quarters of the box contain a sheath plasma  that can be stationary or moving with velocity $v_{\rm sheath}$  \cite{nishi13a,nishi13b,nishi14a,nishi14b}.
This setup is similar to that in our RMHD simulations (\cite{mizuno07a}) 
 that used a cylindrical jet core. Overall, this structure is similar in spirit, although not in scale to that proposed for AGN relativistic jet cores surrounded by a  slower moving sheath, and is also relevant to GRB jets.  However,  here we represent the jet core and sheath as plasma slabs. Initially, the system is charge and current neutral. 

The simulations were performed using a numerical grid with dimension $(L_{\rm x}, L_{\rm y}, L_{\rm z}) = (1005\Delta, 205\Delta, 205\Delta)$, where $\Delta $ is the cell size, and periodic boundary conditions in all directions. The jet and sheath (electron) plasma number density measured in the simulation frame is $n_{\rm jt}= n_{\rm am} = 8$.  The electron skin depth, $\lambda_{\rm s} = c/\omega_{\rm pe} = 12.2\Delta$, where $\omega_{\rm pe} = (e^{2}n_{\rm am}/\epsilon_0 m_{\rm e})^{1/2}$ is the electron plasma frequency and the electron Debye length for the ambient electrons $\lambda_{\rm D}$ is  $1.2\Delta$.  
The jet-electron thermal velocity is $v_{\rm jt,th,e} = 0.014c$ in the jet reference frame, where $c$ is the speed of light.  The electron thermal velocity in the ambient plasma is $v_{\rm am,th,e} = 0.03c$, and ion thermal velocities are smaller by $(m_{\rm i}/m_{\rm e})^{1/2}$. Simulations were performed using an electron-positron ($e^{\pm}$) plasma or an electron-proton ($e^{-}$- $p^{+}$ with $m_{\rm p}/m_{\rm e} = 1836$) plasma for jet Lorentz factors of 1.5, 5.0, and 15.0 with the sheath plasma at rest ($v_{\rm sheath}= 0$) \cite{nishi14a}.
\vspace{-0.0cm}

\vspace{-0.1cm}
The development of the velocity shear surfaces is shown in Figure~\ref{setup}b for $e^{-}$- $p^{+}$ and Figure~\ref{setup}c for $e^{\pm}$ plasmas with $v_{\rm core} = 0.9978c ~(\gamma_{\rm jt}=15)$. For the $e^{-}$- $p^{+}$ case, a nearly DC magnetic field is generated at the shear-surfaces with negative (blue) $B_{\rm y}$ at $z=150\Delta$ and positive (red) $B_{\rm y}$ at $z = 50\Delta$. Additionally, a $B_{\rm z}$ (and $B_{\rm x}$) magnetic field component, shown by the small arrows in Figure \ref{setup}b, is generated at the shear surfaces by current filaments. On the other hand, for the $e^{\pm}$ case a relatively long wavelength ($\sim 100\Delta$) AC magnetic field is generated at the shear surfaces. Note the alternating $B_{\rm y}>0$ (red) and $B_{\rm y}<0$ (blue) along the flow direction. While our results are similar to those found by \cite{liang13a,liang13b}, there are significant important structural differences because their simulations were two-dimensional and used a counter-streaming setup. 

\vspace{-0.5cm}
\subsection{Cylindrical Jet Velocity Shear}
\vspace{-0.35cm}

\begin{figure}[h!]
\includegraphics[width=65mm]{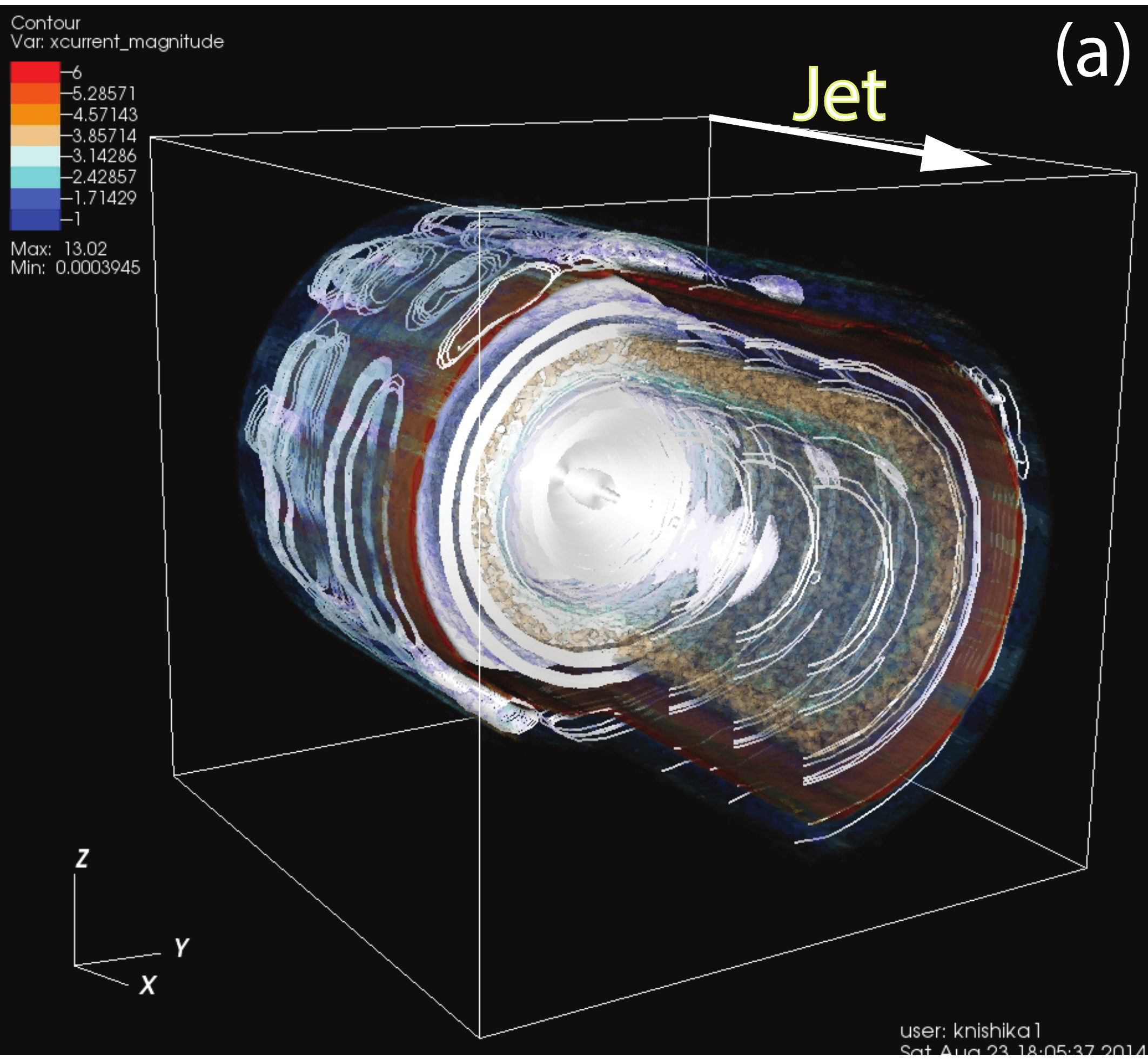}
\vspace{0.1cm}
\includegraphics[width=65mm]{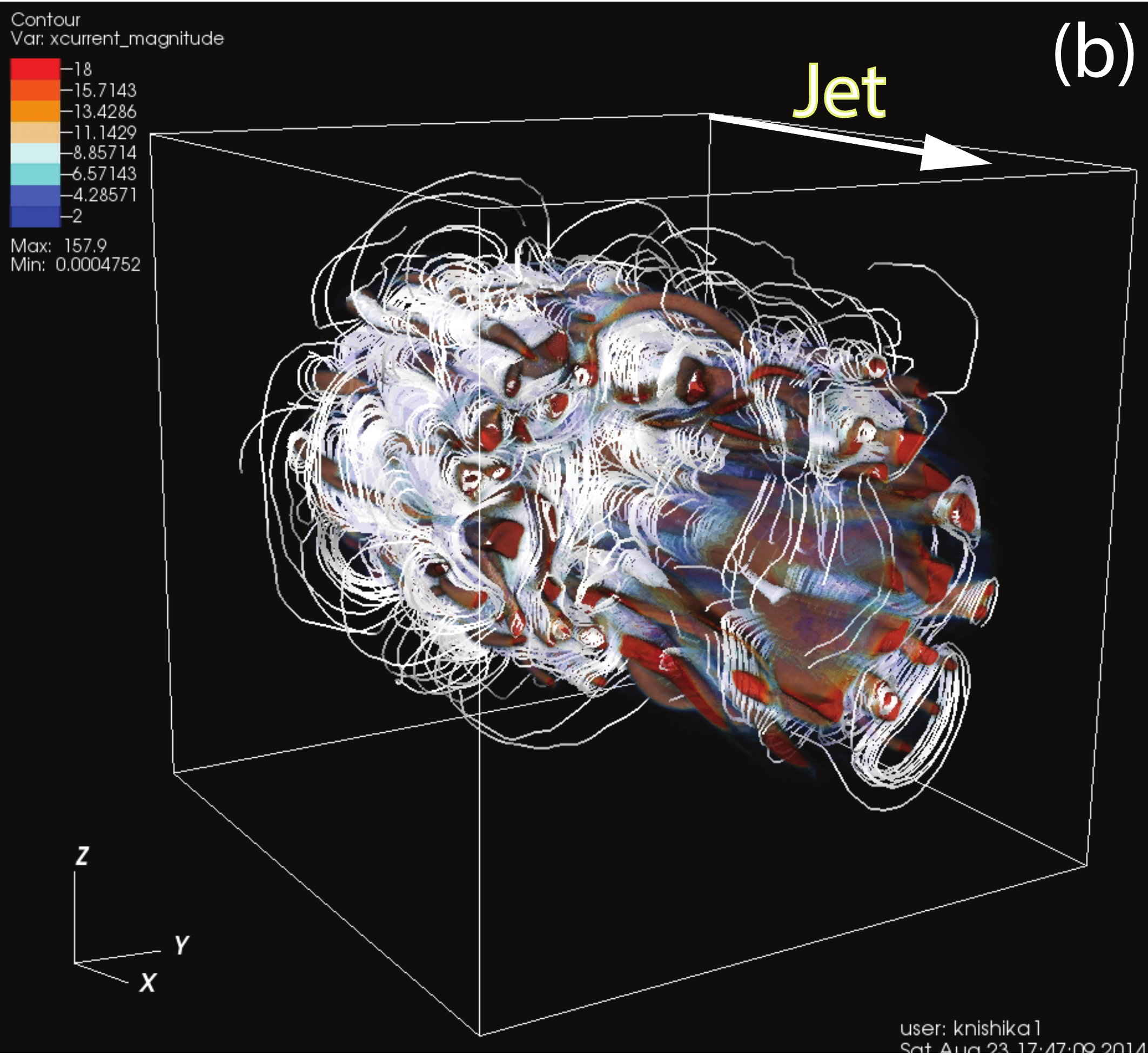}
\vspace{-.2cm}
\caption{\footnotesize \baselineskip 10pt {Isocontour plots of the $J_{\rm x}$ magnitude with magnetic filed lines (one fifth of the jet size)  for (a) an $e^{-}$- $p^{+}$ and (b) an $e^{\pm}$ jet  at simulation time $t =  300\Omega_{\rm pe}^{-1}$. The 3D displays are clipped along the jet and perpendicular to the jet in order to view the interior.}}
\label{fig4}
\vspace{-0.2cm}
\end{figure}

Since relativistic jets and internal filamentary structures are more suitably modeled as intrinsically cylindrical, we have investigated velocity shear in cylindrical geometry for a pair ($e^{\pm}$) and an electron-proton ($e^{-}$- $p^{+}$) jet.   Figure \ref{fig4} shows isocontour images of the $x$ component of the current along with magnetic field lines generated by the kKHI for $e^{\pm}$ and $e^{-}$- $p^{+}$ jets.
The isocontour images show that in the  $e^{-}$- $p^{+}$ jet case currents are generated in sheet like layers and the magnetic fields are wrapped around the jet. On the other hand, in the  $e^{\pm}$ jet case many distinct current filaments are generated near the velocity shear and the individual current filaments are wrapped by the magnetic field. The clear difference in the magnetic field structure between these two cases may make it possible to distinguish different jet compositions via differences in circular and linear polarization.  

\vspace{-0.4cm}
\section{A Combined Shock and Velocity Shear Simulation}
\vspace{-0.2cm}

We have begun ``global" simulations involving injection of a cylindrical jet into an ambient plasma in order to investigate shock (Weibel instability) and velocity shear (kKHI) simultaneously. Previously these two processes have been investigated separately.  In reality a jet or internal filament is injected into an ambient plasma resulting in velocity shear and shocks in a potentially  complicated shock/shear system. 

In order to begin investigation of the combined processes we have performed  a simulation where a relativistic cylincrical jet is  injected  into  an ambient plasma. We used a small system size of $(L_{\rm x}, L_{\rm y}, L_{\rm z}) = (1005\Delta, 131\Delta, 131\Delta$) with jet radius $r_{\rm j} =  20\Delta$ and Lorentz factor $\gamma_{\rm jt} = 5$ to examine the fundamental differences between $e^{\pm}$ and $e^{-}$- $p^{+}$ jet cases and to test synthetic spectra computations. Previous synthetic spectra computations can be found in \cite{nishi11,nishi12a,nishi12b,nishi12c,nishi13a,nishi13b}.

\vspace{-0.5cm}
\subsection{2D Density, Current and Magnetic Field Structure}
\vspace{-0.35cm}

\begin{figure}[h!]
\begin{center}
\includegraphics[width=80mm]{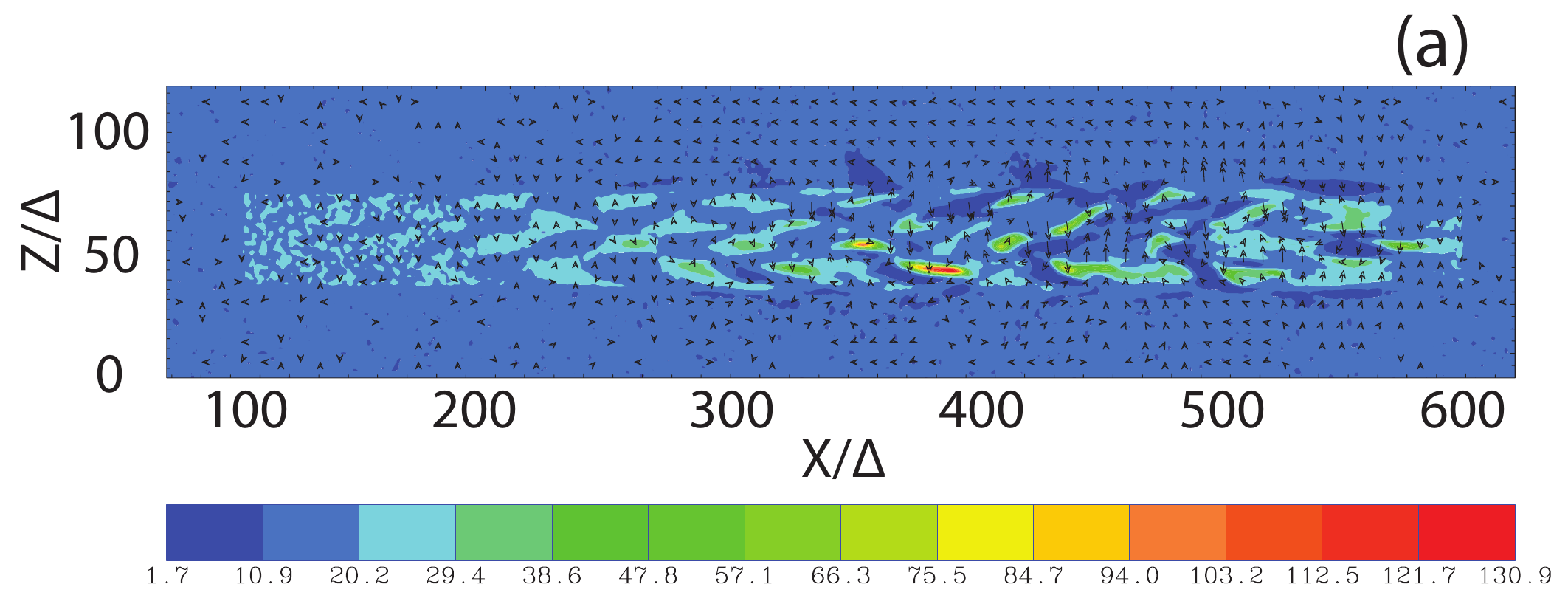}
\end{center}
\begin{center}
\vspace*{-0.5cm}
\includegraphics[width=80mm]{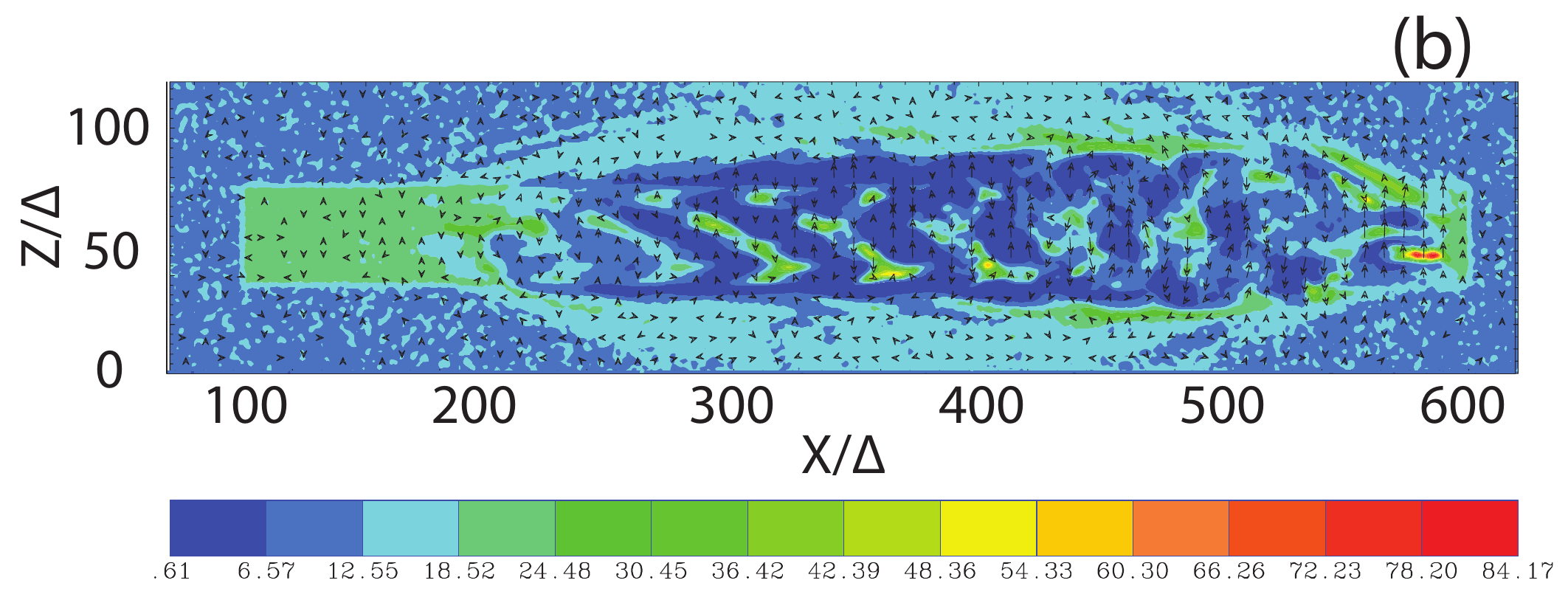}
\end{center}
\vspace*{-0.6cm}
\caption{\footnotesize \baselineskip 10pt  Mid-plane slices of the electron density for jet Lorentz factor $\gamma_{\rm jt} = 5$ at simulation time $t = 500\,\omega_{\rm pe}^{-1}$. The jet is injected at $x/\Delta = 100$, propagates to the right, and the jet front is located at $x/\Delta = 600$. The upper panel (a) shows the electron density structure for the mass ratio $m_{\rm i}/m_{\rm e} = 1836$, and the lower panel (b) for the  electron-positron case.  Associated current structures are  shown in Figure \ref{jxByzsh}.
\label{edBxz}}
\vspace{-0.2cm}
\end{figure}

Figure \ref{edBxz} shows  2D mid-plane slices of the electron density and the transverse magnetic field. 
Current filaments at the jet front are excited by fast current-driven instability in the shock precursor  \cite{lemoi14a,lemoi14b}.
\vspace{-0.1cm}
\begin{figure}[h!]
\begin{center}
\includegraphics[width=40mm]{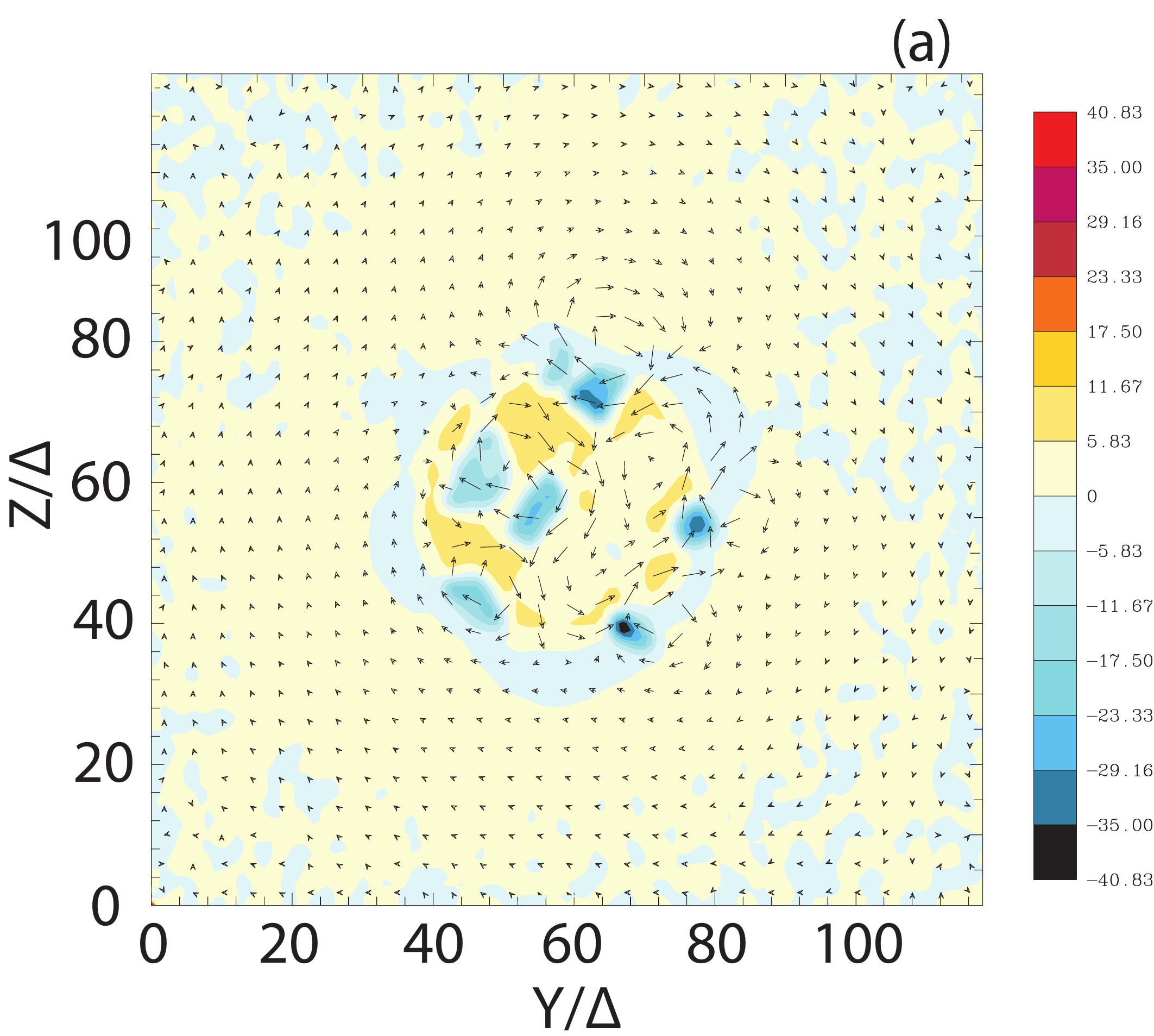}
\hspace*{-0.1cm}
\includegraphics[width=40mm]{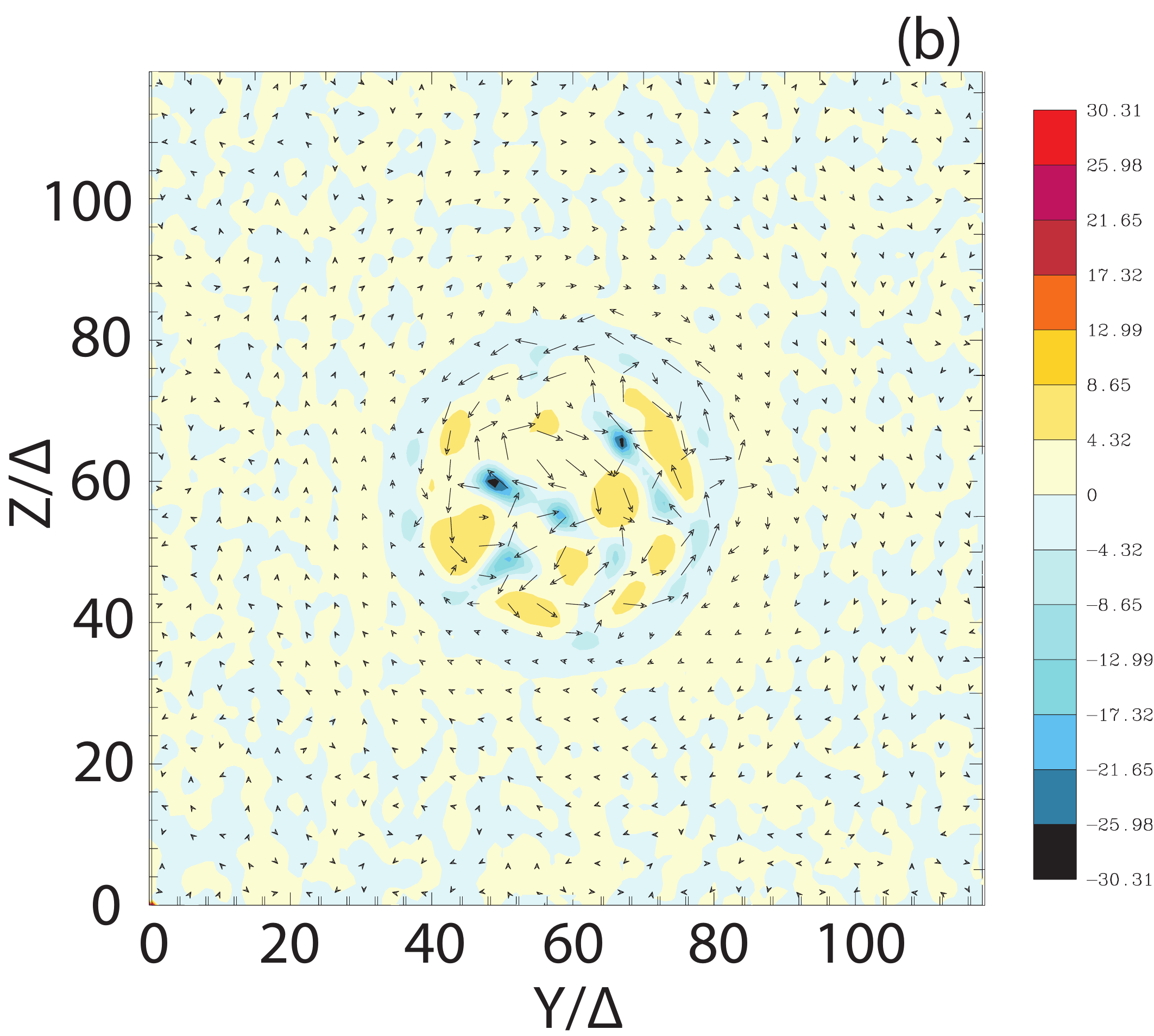}
\end{center}
\begin{center}
\vspace*{-0.5cm}
\includegraphics[width=40mm]{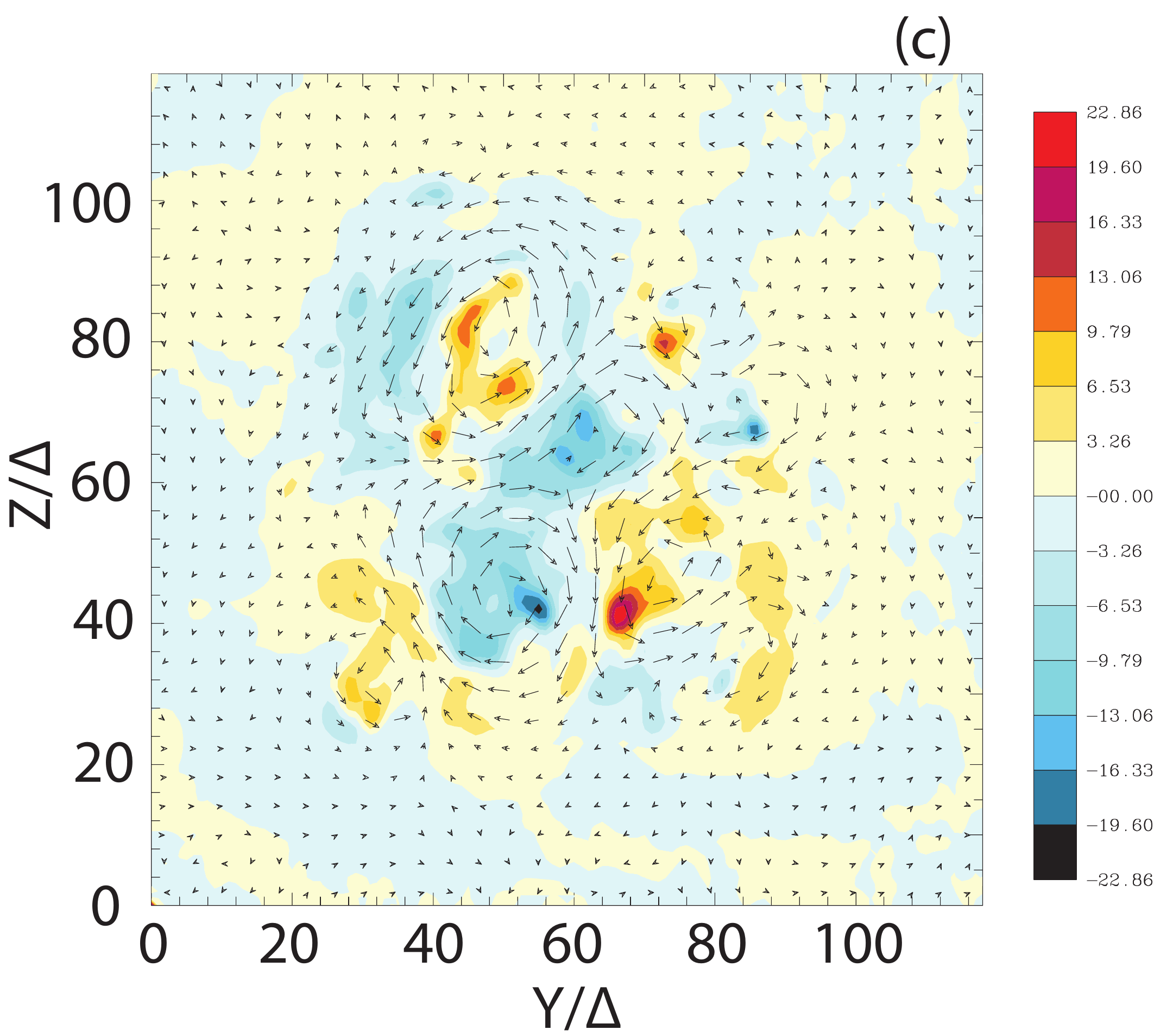}
\hspace*{-0.1cm}
\includegraphics[width=40mm]{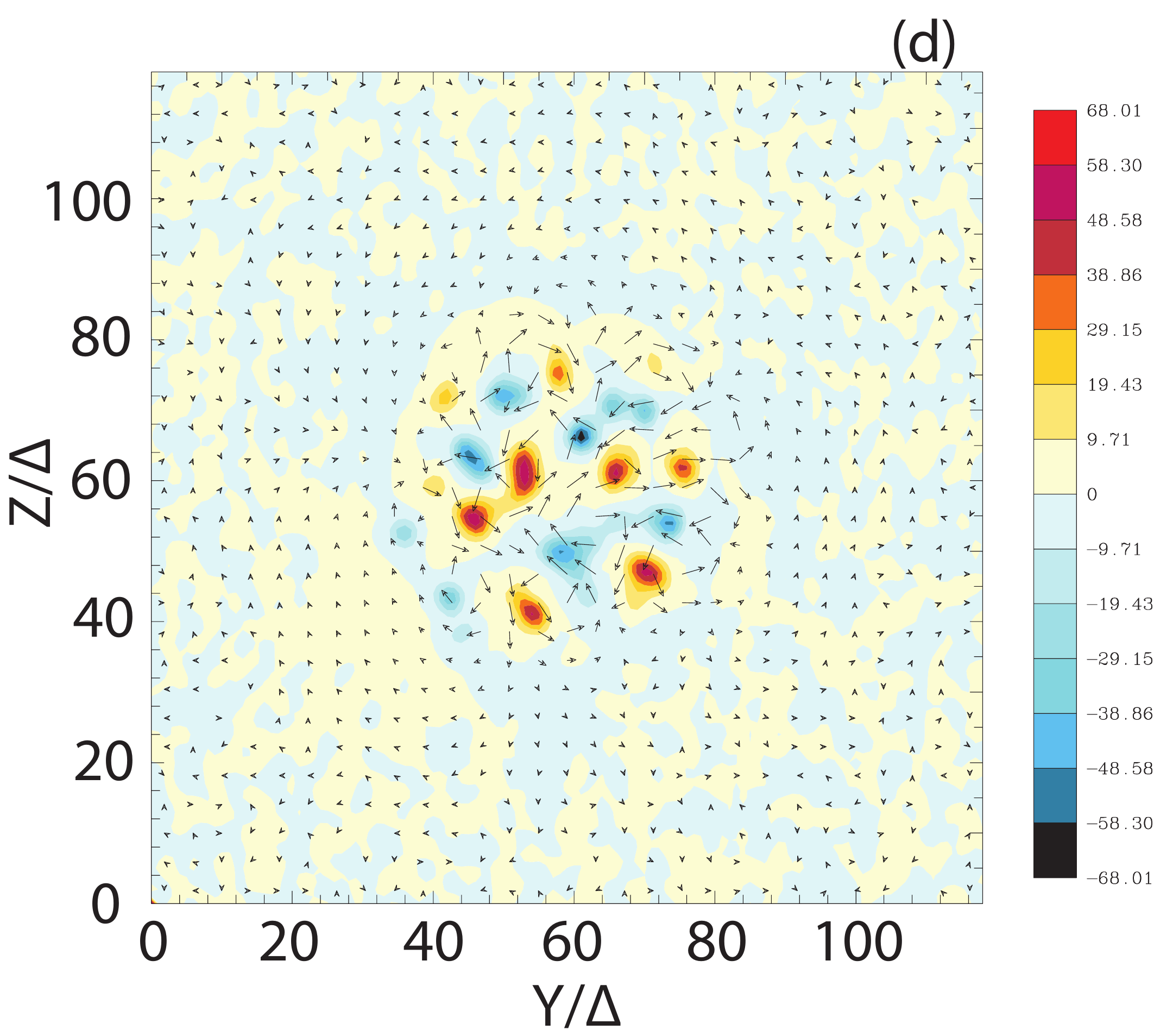}
\end{center}
\vspace*{-0.5cm}
\caption{\footnotesize \baselineskip 10pt  Slices at $x/\Delta = 480$  (left column) and at  $x/\Delta = 580$ (right column) at simulation time $t = 500\,\omega_{\rm pe}^{-1}$ showing curents and magnetic fields for ((a) and (b)) mass ratio $m_{\rm i}/m_{\rm e} = 1836$  and for ((c) and (d)) the electron-positron case. Jets come out of the page at plane center. \label{jxByzsh}}
\vspace{-0.25cm}
\end{figure}
Comparing the electron-proton and electron-positron cases reveals significant electron density structure differences. Jet electrons and protons remain within the jet in the electron-proton case, but  are found outside the jet in the electron-positron case due to mixing with ambient electrons and positrons as the positrons have more mobility than heavy protons.

Our previous simulations of the Weibel instability (e.g., \cite{nishi09a}), 
showed current filaments associated with the growing instability and Figure \ref{jxByzsh} shows the structure of current filaments in cross sections of the jet at two locations. For the electron-proton case the current filaments lie within the jet and a negative current is dominant outside the jet (as in a previous simulation shown in Fig.\ 7a and 7b in Nishikawa et al.\cite{nishi14a}. However, for the electron-positron case, current filaments are found both inside and outside the jet.  In particular, large current filaments can be seen at $x/\Delta = 480$ outside the jet, similar to what was observed in the slab model (Nishikawa et al.\cite{nishi13a,nishi13b,nishi14a,nishi14b}). 

\subsection{3D Current and Magnetic Field Structure}
\vspace{-0.35cm}

\begin{figure}[!h]
\includegraphics[width=80mm]{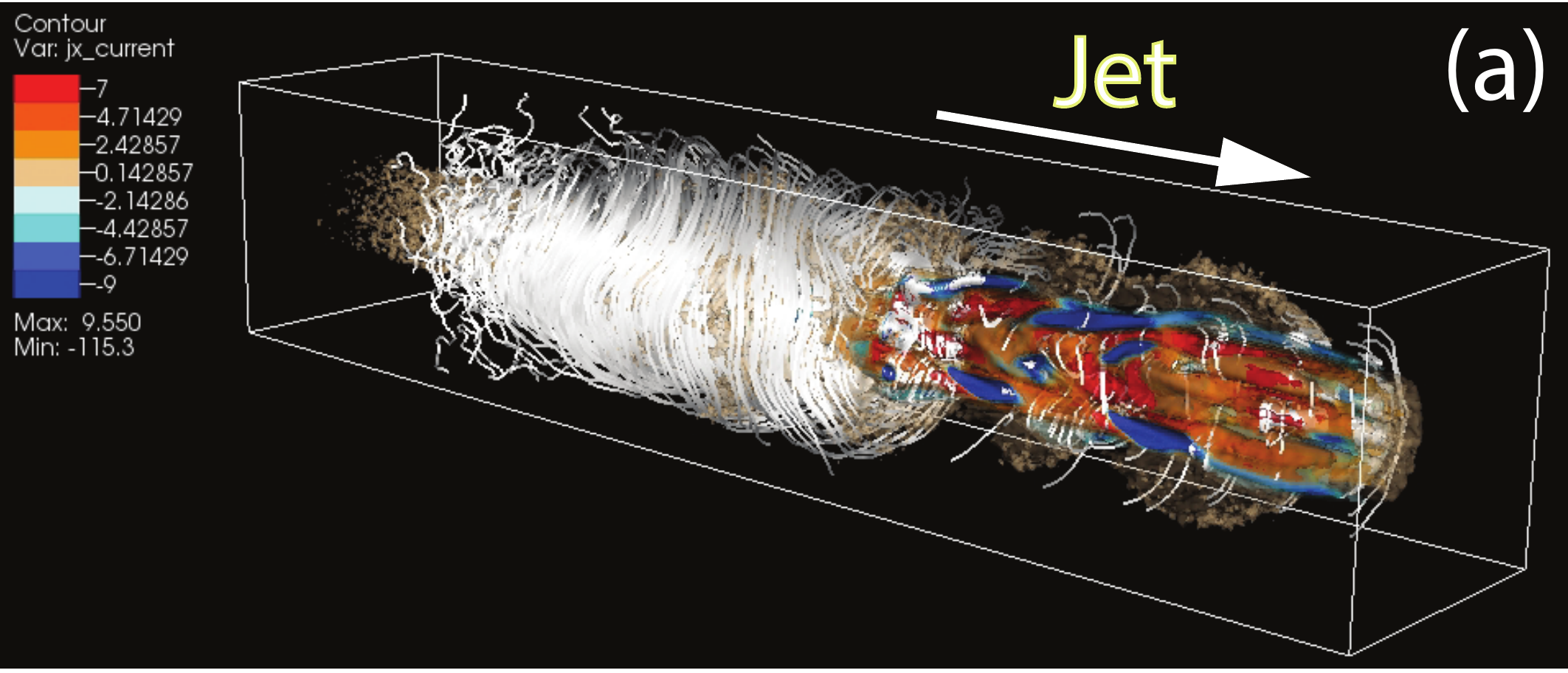}
\includegraphics[width=80mm]{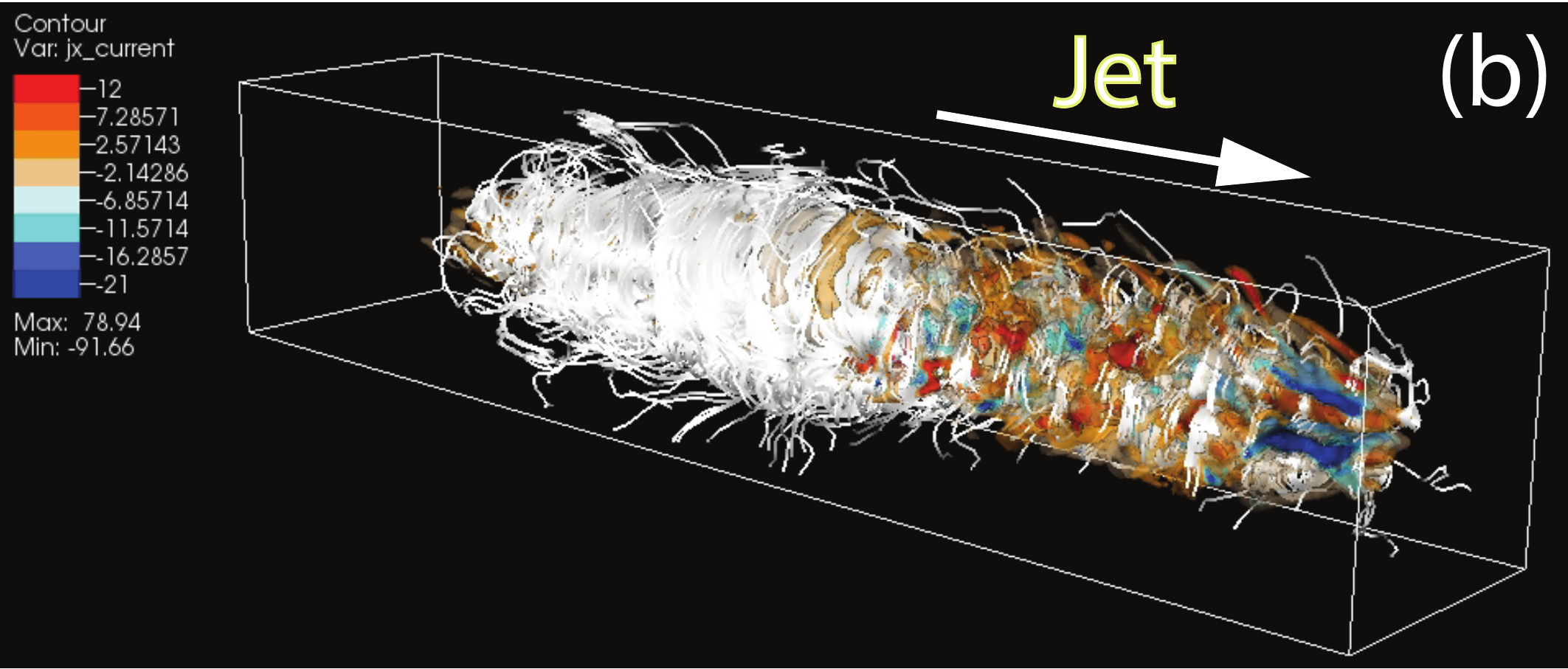} 
\vspace*{-0.1cm}
\caption{\footnotesize \baselineskip 10pt Global jet simulations for $e^{-}$- $p^{+}$ and $e^{\pm}$ jets at time $t =  500\Omega_{\rm pe}^{-1}$ . Panel (a) shows the $e^{-}$- $p^{+}$ jet where long and continuous current filaments are confined within the jet. Panel (b) shows the  $e^{\pm}$ jet where short current filaments are found within and outside the jet.
 \label{3DJxB}}
\vspace{-0.2cm}
\end{figure}

Figure \ref{3DJxB} shows 3D current filament isosurfaces along with magnetic field lines that are generated by the Weibel instability and by the kKHI. Only the front part of the jet is displayed ($120 < x/\Delta < 520$) and the jet is propagating from back left to forward right. The cube is
\vspace*{-0.0cm}
clipped at $x/\Delta  = 320$ and $y/\Delta = 66$ in order to view cross sections parallel and perpendicular to the jet axis. For both cases, compact  current filaments  are confined mostly within the jet at the jet front (see also Figure \ref{jxByzsh}b and \ref{jxByzsh}d). The magnetic fields wrap around the current filaments. 
Since the simulation system is small, current structures are at an early stage of development, and a fully developed shock system is not yet formed at the jet front. In the $e^{-}$- $p^{+}$ case long and continuous current filaments remain confined within the jet along the velocity shear surface behind the jet front(Fig. \ref{3DJxB}a). For the  $e^{\pm}$ case short current filaments are found within and outside the jet along the velocity shear surface behind the jet front (Fig. \ref{3DJxB}b). This presence of current filaments inside and outside the jet in the $e^{\pm}$ case is also observed in slab geometry (see Fig.\ \ref{setup}a and Nishikawa et al.\cite{nishi14a}), where current filaments were found farther outside the jet at smaller jet Lorentz factors. 

Small system simulations cannot fully distinguish between current filaments generated by the Weibel instability and those generated by the kKHI, and velocity shear effects are insufficiently resolved. Larger simulations need to be performed to clearly reveal the differences between the two cases, e.g., Nishikawa et al.\cite{nishi09a} and Choi et al.\cite{choi14}, and allow proper evaluation of emission from the jet boundary.

\vspace{-0.55cm}
\section{Future Work: Helical Magnetic fields and Reconnection}
\vspace{-0.35cm}

Our preliminary 3D global structure simulations indicate the importance of using global simulations to investigate the combined evolution of collisionless shocks and velocity shears.  In order to resolve collisionless shock and velocity shear structures we will perfume global jet simulations using much larger systems with $(L_{\rm x}, L_{\rm y}, L_{\rm z}) = (8005\Delta, 405\Delta, 405\Delta$).  Additionally, a very large system is needed in order to scale kinetic processes to larger scale jet structures. A larger system will allow us to obtain a much more reasonable picture of the evolution of the magnetic field and subsequently the dependence of spectra and polarization on jet composition through synthetic emission computations. 

Recently, Markidis et al.\cite{marki14} performed three-dimensional PIC simulations of a flux rope instability using a single flux rope with a simple screwpinch configuration \cite{freid07}, i.e.,  the helical magnetic field has a constant pitch (e.g., \cite{mizunno12}).  An artificial ion to electron mass ratio equal to 25 was chosen to reduce the simulation execution time, and the initial current was carried by the electrons (ions were initially stationary). The simulation revealed magnetic reconnection during the kink instability of the flux rope. Secondary signatures of magnetic reconnection included a quadrupolar structure in the density, a bipolar structure in the Hall field, and a reconnection associated electric field in proximity to the reconnection region. 

In our future work we will inject jets like those shown in Figure \ref{3DJxB} but containing a helical magnetic field like that implemented in Markidis et al.
\cite{marki14} and using a setup like that used in Wieland et al. \cite{wieland14}. This setup avoids transient phenomena due to an infinitely sharp contact discontinuity at the colliding front and avoids an artificial magnetic field pileup. In our setup we will generate a helical magnetic field via  faster jet ions (protons or positrons) instead of electrons. This configuration will allow investigation of (1) the effect of helical magnetic field on growth of the Weibel instability and the kKHI, (2) the possible development of MHD-like kink and/or global KHI, and (3) the development of magnetic reconnection.

%




\vspace*{-0.6cm}
\section{Conclusions}
\vspace*{-0.3cm}

We have presented 3D PIC simulations of  collisionless shock and velocity shear development mediated by the Weibel instability  and the kKHI for both electron-positron and electron-ion plasmas. The processes studied here are important in AGN and GRB jets that are expected to have shocks and velocity shears between faster and slower moving plasmas both within the jet and at the jet external medium interface. 


We have shown via shock simulations that
shock structure depends on the composition of the plasma,  e.g., Choi et al.\cite{choi14} and Nishikawa et al.\cite{nishi09a}.
%
The collisionless electron-ion ($m_{\rm i}/m_{\rm e} = 16$) shock shows
a sharper rise in the electron density at the forward shock than the
electron-positron case (see Fig.\ref{weib}a and compare to Fig.\ 1a in Nishikawa
et al.\cite{nishi09a}).  This sharper rise occurs because in the
electron-positron case jet electrons propagate through the forward shock
to the jet front but in the electron-ion case the jet electron density
declines in front of the forward shock.  This decline in the
electron-ion case is due to the ambipolar electric fields created at the
jet front by the heavier ions.

We have shown via velocity
shear simulations that velocity shear structure depends on the
composition of the plasma and the jet Lorentz factor, e.g., Nishikawa et
al.\cite{nishi13b,nishi14a,nishi14b}.
The growth rate for the kKHI for the mildly relativistic jet case ($\gamma_{\rm j} = 1.5$) is larger than the relativistic jet case ($\gamma_{\rm j} = 15$).
In particular, the different magnetic field velocity shear structure associated with electron-positron composition versus electron-proton composition should have consequences for the polarization of jets in very 
high-resolution radio imaging. For a simple cylindrical geometry velocity shear case an electron-proton jet  primarily builds magnetic field in the toroidal direction at the velocity shear surface.  In contrast, a pair-plasma jet  generates sizable radial field components that are only about a factor of two weaker than the toroidal field. In either case, strong electric and magnetic fields in the velocity shear zone will also be conducive to particle acceleration. 

When global jet injection simulations are performed the combination of shock and velocity shear structures makes the situation more complicated but our preliminary simulations show clear differences  between electron-positron and electron-proton plasmas. Our preliminary simulations are too short for definitive statements on the efficacy of the process and the resulting spectra. However, it is clear that the magnetic field structure along with particle acceleration and transport in compact regions will be necessary for a realistic assessment and interpretation of observed emission spectra and polarization.

\vspace{-0.55cm}
\begin{acknowledgments}
\vspace{-0.35cm}

K.-I. Nishikawa and/or P.\ Hardee have been supported by NSF awards AST-0908010 and AST-0908040, and by NASA awards NNG05GK73G, NNX07AJ88G, NNX08AG83G, NNX08AL39G, NNX09AD16G, NNX12AH06G, NNX13A P-21G, and NNX13AP14G. J.\ Niemiec  has been supported by the Polish National Science Centre via  awards DEC-2011/01/B/ST9/03183 and DEC-2012/04/A/ST9/00083. Y.\ Mizuno has been supported by the Ministry of Science and Technology of Taiwan via awards NSC 100-2112-M-007-022-MY3 and MOST 103-2112-M-007-023-MY3. M.\ Pohl is supported via award PO 1508/1-2 from the Deutsche
Forschungsgemeinschaft. Simulations were performed on Columbia and Pleiades at the NASA Advanced Supercomputing (NAS), Kraken and Nautilus at The National Institute for Computational Sciences (NICS), and Stampede at The Texas Advanced Computing Center. This research was begun during the program ``Chirps, Mergers and Explosions: The Final Moments of Coalescing Compact Binaries'' at the Kavli Institute for Theoretical Physics, which is supported by the NSF via grant PHY05-51164. Initial kKHI work began
at the Aspen Center for Physics workshop  ``Astrophysical Mechanisms of Particle Acceleration and
Escape from the Accelerators''  (September 1-15, 2013).
\end{acknowledgments}


\end{document}